# Can structure influence hydrovoltaic energy generation? Insights from metallic 1T' and semiconducting 2H phases of MoS$_2$

Kaushik Suvigya[1], Saini Lalita[1], Siva Nemala Sankar[3], Andrea Capasso[3], Li-Hsien Yeh[4,5,6] and Kalon Gopinadhan[1,2]

[1]Department of Physics, Indian Institute of Technology Gandhinagar, Gujarat, India- 382355

[2]Department of Materials Engineering, Indian Institute of Technology Gandhinagar, Gujarat, India- 382355

[3]International Iberian Nanotechnology Laboratory 4715-330 Braga, Portugal

[4]Department of Chemical Engineering, National Taiwan University of Science and Technology, Taipei 10607, Taiwan

[5]Graduate Institute of Energy and Sustainability Technology, National Taiwan University of Science and Technology, Taipei 10607, Taiwan

[6]Advanced Manufacturing Research Center, National Taiwan University of Science and Technology, Taipei 10607, Taiwan

**Abstract**

**Hydrovoltaic power generation from liquid water and ambient moisture has attracted considerable research efforts. However, there is still limited consensus on the optimal material properties required to maximize the power output. Here, we use laminates of two different phases of layered MoS$_2$ – metallic 1T' and semiconducting 2H – as representative systems to investigate the critical influence of specific characteristics, such as hydrophilicity, interlayer channels, and structure, on the hydrovoltaic performance. The metallic 1T' phase was synthesized via a chemical exfoliation process and assembled into laminates, which can then be converted to the semiconducting 2H phase by thermal annealing. Under liquid water conditions, the 1T' laminates (having a channel size of ~6 Å) achieved a peak power density of 2.0 mW.m$^{-2}$, significantly outperforming the 2H phase (lacking defined channels) that produced 2.4 µW.m$^{-2}$. Our theoretical analysis suggests that energy generation in these hydrophilic materials primarily arises from electro-kinetic and surface diffusion mechanisms. These findings highlight the crucial role of phase-engineered MoS$_2$ and underscore the potential of 2D material laminates in advancing hydrovoltaic energy technologies.**

## 1. Introduction

A green and promising direction to meet the rising global energy demand could be the utilization of ambient moisture to produce electricity without the consumption of an external mechanical input[1]. Moreover, water evaporation is a natural and spontaneous phenomenon, and the energy harnessed from this is clean and safe. Two-dimensional (2D) materials have gained significant interest in water evaporation-based electricity generators (EEGs) because (i) they possess (negatively) charged surfaces, which can lead to electro-kinetic effects upon interaction with water[2], and (ii) their interlayer spacing allows for the formation and overlap of electric double layer (EDL). The EDL overlap occurs when a fluid like water is confined in nanochannels with a diameter less than its Debye length. Thus, quick capillary action can harvest energy from the uninterrupted water flow inside these layers. The early materials in water energy harvesting primarily included carbon nanostructures such as carbon black[1,3], graphene, and its derivatives[4,5]. These materials continued to be explored, along with several others that soon entered the picture, such as MOFs[6] and biomaterials[7–9] to hydrogels[10–12] and different hybrid composites[5,13,14]. Broadly, the active materials for efficient water energy harvesting possessed specific properties, such as the presence of functional groups (like carboxyl, carbonyl, hydroxyl, etc.), large specific surface area, hydrophilicity, and electrical conductivity. While the focus remained predominantly on producing higher output, there is no consensus on the underlying mechanisms and how factors such as conductivity or structure affect the process of energy generation. The materials in which these properties can be effectively tuned are, therefore, highly beneficial.

Very few works have explored the properties of $MoS_2$ in water energy harvesting, and they are limited by their low output power. For instance, a phase-engineered $MoS_2$ film reported by He *et al.* produced 19 mV and 6.24 μA under moisture at an internal resistance of 3 kΩ[15]. The power density was 42 μW.cm$^{-3}$ at ~10 mV. A chemical vapor deposition (CVD)-grown $MoS_2$ monolayer reported by Aji *et al.* produced an output voltage of ~5 V by the motion of 1 M aqueous NaCl droplet on it[16]. However, the voltage produced in this manner was instantaneous, and the power generated was 7 nW. Further, there is no study comparing the efficacy of water energy harvesting on the different phases of a single material system. Herein, we chose $MoS_2$, which offers two phases with different conductivities: the metallic 1T' phase and the semiconducting 2H phase. The stacking in the 2H phase is BaB AbA (capital letters for S atoms, small letters for Mo atoms), and the coordination of Mo with S is trigonal prismatic. The S atoms in the upper layer are located directly above those in the lower layer[17]. The stacking in the metallic 1T phase is AbC, where upper S atoms are shifted from lower S atoms, and Mo atoms occupy the octahedral holes of the S atoms. Due to different crystal symmetries, the structures of both phases are substantially different. Unlike the 2H phase, the 1T' (distorted 1T) phase of $MoS_2$ remained largely unexplored. Recently, the 1T' phase of $MoS_2$ was prepared by chemical exfoliation by Hu *et al.* to demonstrate pH-dependent switching of water permeability[18]. In this work, we highlight the crucial role of structure and channels by fabricating water evaporation-based electricity generators (EEGs) from the laminates of the two phases. The power density obtained from metallic 1T' laminates was found to be 2.0 mW m$^{-2}$, three orders higher than that of semiconducting 2H laminates, owing to the structure and presence of well-defined channels in the metallic laminates.

## 2. Results and Discussion

We fabricated the MoS$_2$ membranes in two phases (details in the Experimental Section). In brief, the metallic MoS$_2$ phase was prepared by lithium intercalation into the 2H phase bulk MoS$_2$ powder in an argon atmosphere (Fig. 1a). Upon lithium-ion intercalation, Mo coordination changes from trigonal prismatic (H) to octahedral (T), and a first-order phase transition occurs. This chemical exfoliation of MoS$_2$ by lithium intercalation results in electron transfer from metal ions to the lowest-lying unoccupied energy levels of the metallic d-bands of the MoS$_2$ crystal[15]. The original crystal lattice becomes unstable due to the charge transfer which changes the electron configuration from d$^2$ to d$^3$ (Ref.[19]). This results in a negative surface charge on it. This chemically exfoliated structure, represented as 1T', is distorted from the ideal T structure, and has Mo atoms arranged into zigzag chains[17]. It is the energetically preferred metastable phase and is marked by the presence of exchangeable cations due to excess negative charge. The semiconducting 2H phase was prepared by annealing the metallic phase in a vacuum oven. It is a stable phase with hexagonal symmetry.

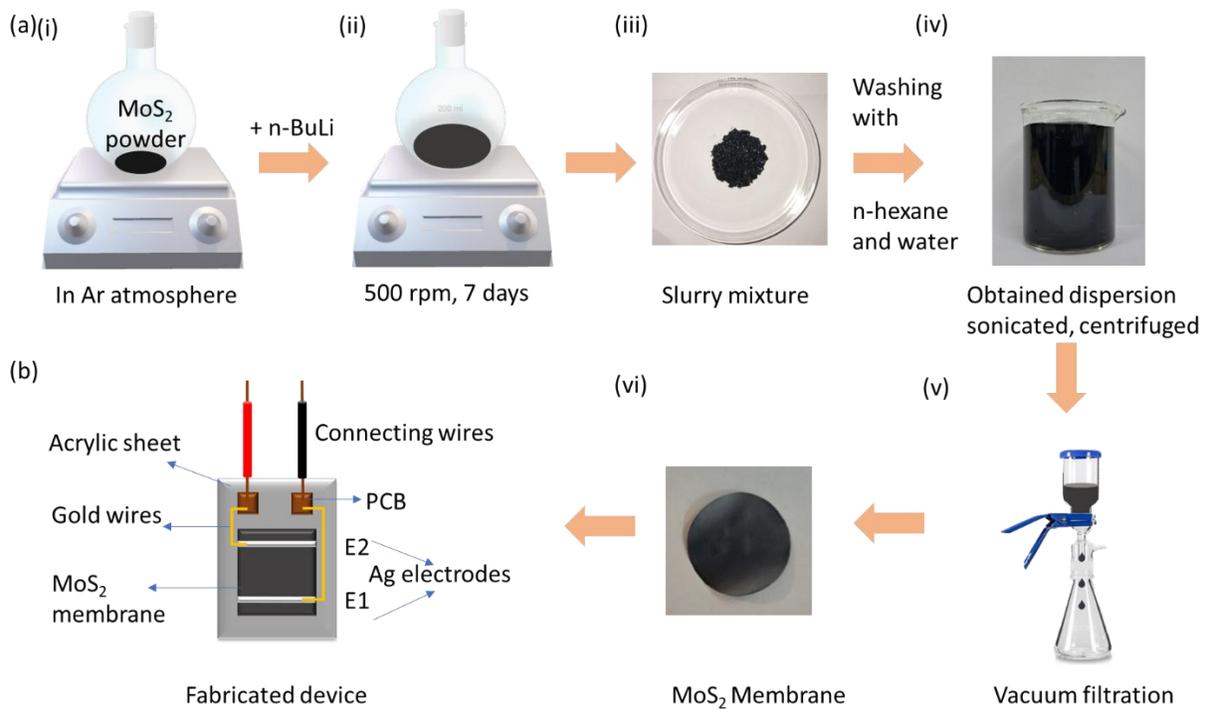

Fig. 1. Synthesis of MoS$_2$ 1T' phase and EEG device fabrication. (a) (i-vi) Steps of synthesis of 1T' phase by introducing n-BuLi to the bulk MoS$_2$ powder in argon atmosphere. (b) Schematic of a water evaporation-based electricity generator (EEG).

The laminates of both phases were characterized by various characterization techniques. Scanning electron microscopy (SEM) imaging revealed a highly lamellar structure of the 1T' membrane (Fig. 2a). The SEM image of the 2H phase MoS$_2$ membrane is shown in Fig. S1. A membrane thickness of 1 μm was determined by the cross-sectional SEM imaging. X-ray diffraction (XRD) was performed to obtain the interlayer spacing of the 1T' and 2H membranes. As seen in Fig. 2b, the 2H membrane shows the characteristic peak of MoS$_2$ at 2θ = 13.6°, corresponding to the (002) plane with an interlayer spacing of 6.5 Å. On the other hand, 1T' membrane shows two peaks at 2θ = 7.1° (001 plane) and 14.2° (002 plane), indicating an increased interlayer spacing of 12.5 Å. The hydration of MoS$_2$ layers with Li$^+$ ions due to osmotic pressure-driven water intercalation causes swelling. Thus, the larger interlayer space in 1T' is due to the presence of two layers of intercalated water. Although the interlayer spacing (also defined as van der Waals distance, $d_{vdW}$) of 2H MoS$_2$ is 6.5 Å, this whole space is not available for the transport of a species[20] (Supplementary Section S2.1). The actual space available, known as the van der Waals gap ($h_{vdW}$), is estimated by subtracting the thickness of the electron cloud. The compact structure of 2H MoS$_2$ does not allow the transport of any species through its interlayer space (Fig. S2). Thus, we can take 6.5 Å as the upper bound of the thickness of the electron cloud to calculate the $h_{vdW}$ for 1T'. It is estimated to be 6 Å, enough to accommodate two layers of water (thickness of monolayer water = 2.8 Å). A freshly prepared 1T' phase membrane captures excess water inside the interlayers due to the presence of Li$^+$ ions. Thus, when analyzed just after preparation, no peak corresponding to MoS$_2$ is observed (Fig. 2c). However, upon drying the membrane for 10 minutes under an IR lamp (Murphy Infrared Bulb 150 W, 230 V), 1T' phase (001 plane) is observed. During the membrane preparation, it was noticed that a 1T' membrane appears to have a glossy black finish just after preparation as it retains water. It loses this extra water as it dries and slowly turns matte black (insets Fig. 2c). Further, when this 1T' membrane is annealed at 200 °C in an inert atmosphere, the hydrated Li$^+$ inside the interlayers gets removed (as confirmed by the XPS studies, which is discussed later), the interlayer spacing decreases, and the phase conversion from 1T' to 2H takes place.

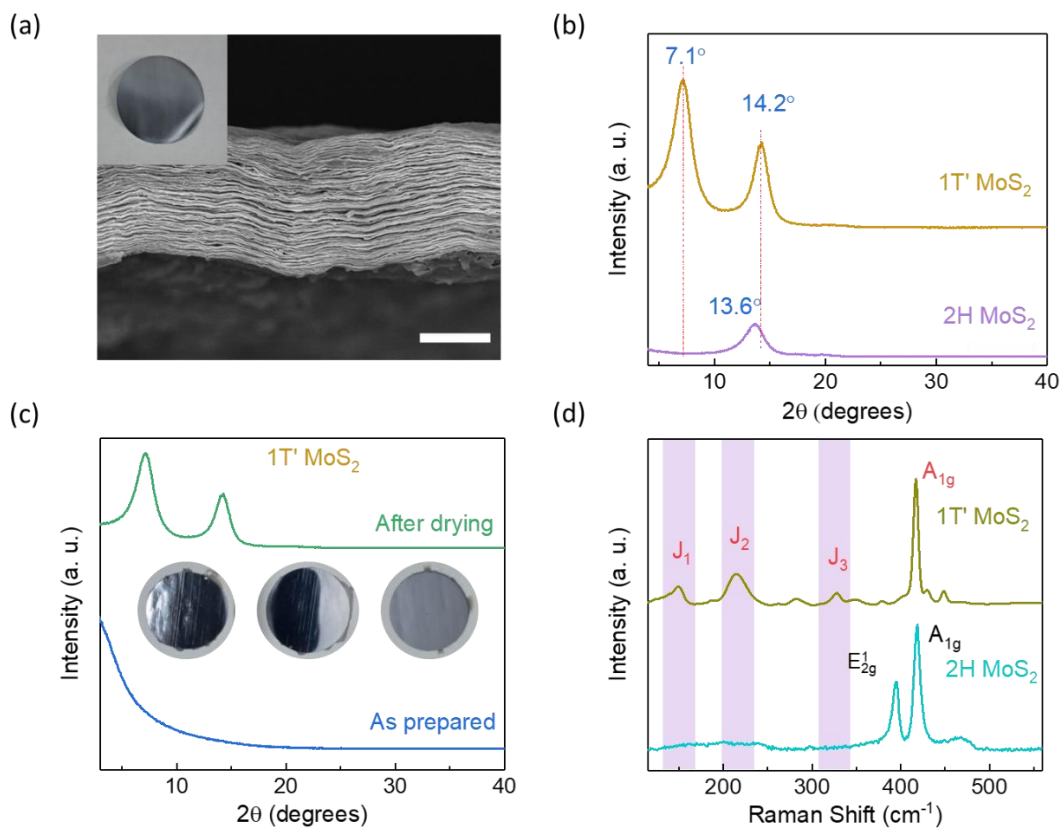

Fig. 2. X-ray diffraction and Raman spectra of 1T' and 2H MoS$_2$ membranes. (a) Cross-sectional SEM image of 1T' MoS$_2$ membrane. Scale bar 500 nm. Inset: camera image of a free-standing 1T' MoS$_2$ membrane. (b) XRD patterns of 1T' and 2H MoS$_2$ membranes. An additional peak at 7.1° confirms that the interlayer spacing for 1T' MoS$_2$ is larger than that of 2H MoS$_2$. (c) Effect of drying the 1T' membrane. Insets from left to right: camera images of 1T' membrane on PVDF substrate – full wet (i.e., as prepared); half wet and half dry; full dry. (d) Raman spectra of 1T' and 2H MoS$_2$ membranes. The J$_1$, J$_2$, and J$_3$ modes are characteristic of the 1T' phase.

The formation of the phases was also confirmed by Raman spectroscopy (Fig. 2d). The Raman spectra of 2H MoS$_2$ contain peaks corresponding to $E^1_{2g}$ and $A_{1g}$ modes, indicating the in-plane and out-of-plane vibrational modes of the MoS$_2$ layer, respectively. Additionally, the 1T' MoS$_2$ has three distinct features at 149, 214, and 327 cm$^{-1}$, corresponding to the J$_1$, J$_2$, and J$_3$ modes[21]. X-ray photoelectron spectroscopy (XPS) was performed to determine the atomic compositions of the two phases. Fig. 3a-d shows the Mo 3d, S 2s, and S 2p regions of the MoS$_2$ membranes in 1T' phase and 2H phase. The 1T' phase has the Mo 3d and S 2s peaks shifted to lower binding energy by ~0.5 eV than the 2H phase peaks[22]. These measurements also reveal the presence of S vacancies: 34.0% in the 1T' phase and 32.5% in the 2H phase (Supplementary Section S2.2) which provide a negative surface charge on the membranes[23,24]. The XPS spectra also confirmed the presence of lithium inside the interlayers of the 1T' MoS$_2$ membrane (Fig. 3e). After annealing the 1T' membrane and its phase conversion to 2H, the removal of hydrated Li$^+$ present inside the interlayers is confirmed by the absence of Li 1s peak in the 2H membrane (Fig. 3e). The ultraviolet-visible (UV-visible) absorption spectra of the two phases is shown in Fig. 3f. The 2H phase has two absorption peaks at 674 nm and 611 nm, originating from A and B direct excitonic transitions, respectively. A weak, broader peak is observed at ~450 nm from the C excitonic transition. However, these features are absent in the UV-visible spectrum of the 1T' phase, in agreement with an earlier report[25].

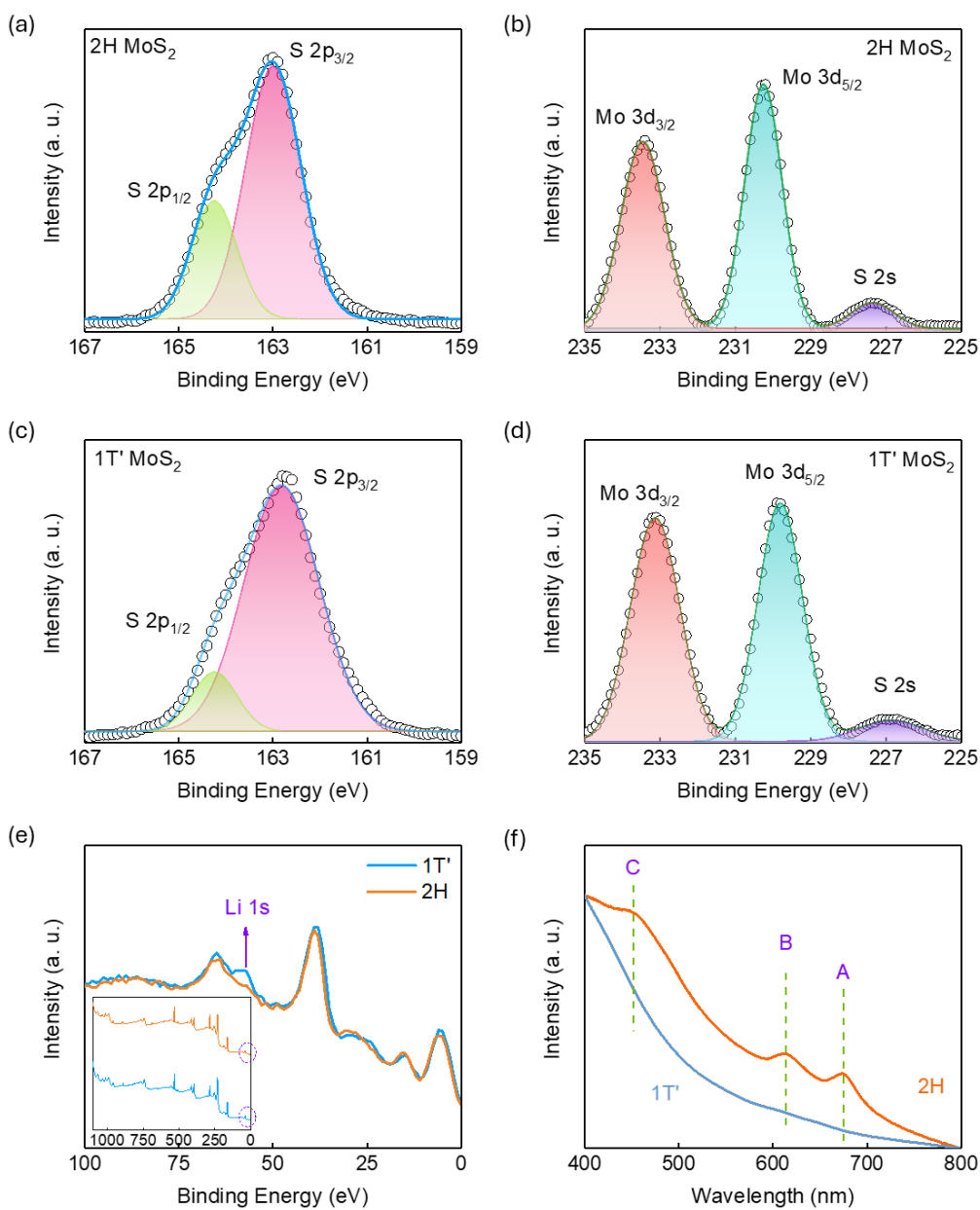

Fig. 3. XPS and Ultraviolet-visible spectra of 1T' and 2H MoS$_2$ membranes. (a) S 2p and (b) Mo 3d peaks in 2H MoS$_2$ membranes. (c) S 2p and (d) Mo 3d peaks in 1T' MoS$_2$ membranes. (e) XPS survey of 1T' MoS$_2$ membranes shows the Li 1s peak, which is absent in the 2H MoS$_2$ membranes. Inset – complete XPS survey. (f) UV-visible spectra of the MoS$_2$ membranes show A, B, and C excitonic peaks in the 2H phase, while it is absent in the 1T' phase. The blue color corresponds to the 1T' phase; the orange color corresponds to the 2H phase.

After the characterizations and phase confirmations, we fabricated the EEGs (Fig. 1b) from both membranes for electricity generation, as described in the Experimental Section. The background short-circuit current (at 0 V) and open-circuit voltage (at 0 A) were monitored with time in ambient moisture conditions. For the 1T' device of width 11 mm, we obtained a background current and voltage on the order of $10^{-8}$ A and $10^{-5}$ V, respectively (Fig. S3a-b). Using the cross-sectional area of $1.1 \times 10^{-8}$ m$^2$, the power density was calculated to be 0.1 mW.m$^{-2}$ (Fig. 4a). Similarly, for the 2H device of membrane width 7 mm, we obtained a background current and voltage on the order of $10^{-12}$ A and $10^{-6}$ V, respectively (Fig. S3c-d). Due to the absence of well-defined channels in 2H phase membranes, the effective area is the area between electrodes E1 and E2, which leads to a very low power density of $2.4 \times 10^{-10}$ mW.m$^{-2}$ (0.2 pW m$^{-2}$) (Fig. 4b). Note that these are the powers generated from the moisture in the ambient environment laboratory conditions (25 °C, 56% RH). Moreover, the background current of 1T' device is four orders larger than that of 2H device, owing to the conductive 2D metallic 1T' MoS$_2$, which significantly reduces the membrane resistance and thus enhances the current[26].

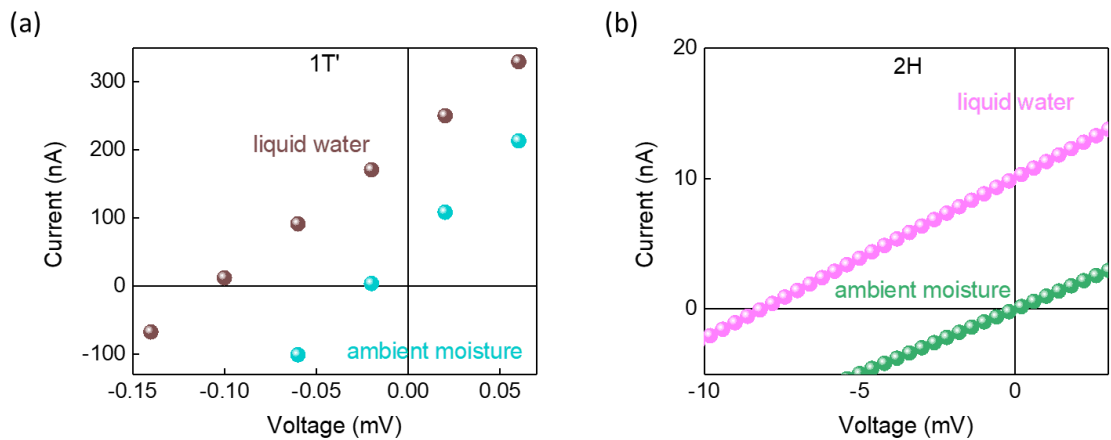

Fig. 4. Performance of 1T' and 2H EEGs. (a) Current vs. voltage characteristics of 1T' device. (b) Current vs. voltage characteristics of the 2H device. The applied voltage was varied from -0.2 V to +0.2 V with a step of 20 mV. The x-axis scale has been zoomed in for clear representation of short-circuit current and open-circuit voltage.

To measure the electricity generated from contact with liquid water, the EEG was dipped into water in a beaker such that E1 was immersed entirely, and the water meniscus was in the middle of E1 and E2. The current (at 0 V) and voltage (at 0 A) were recorded over time for both EEGs. For the 1T' device, we obtained the current and voltage of the order of $10^{-7}$ A and $10^{-4}$ V, respectively (Fig. S3), and the power density was calculated from the current-voltage (I-V) characteristics to be 2.0 mW.m$^{-2}$ (Fig. 4a). The current and voltage outputs are stable at an internal resistance of ~0.5 kΩ, lower than what has been reported previously for MoS$_2$ (Ref.[15]). For the 2H device, the current and voltage in the presence of water were ~$10^{-8}$ A and ~10 mV, respectively. However, the power density was still smaller and around 2.4 µW.m$^{-2}$ (Fig. 4b).

To confirm whether evaporation is the origin of the generated power, we conducted further experiments with a 1T' device (Supplementary Section S2.3). For the first experiment, the EEG was sealed properly using parafilm, and it was observed that the current level gradually dropped by 90%. When the EEG was unsealed again, the current regained its original value (Fig. S4a). As the EEG was sealed, the evaporation rate dropped, and consequently, the current decreased. For the second experiment, air flow was turned on above the EEG device, and it was observed that the current level increased by approximately 12 times (Fig. S4b). This is because the water evaporates faster in the wind flow. Such changes were found to be repeatable and conclusively prove that evaporation drives the current and, hence, the generated power.

What are the possible reasons for the difference in power density generated by the two phases? As liquid-solid interface interactions are influenced by the wettability of the solid surface by the liquid[27], it is essential to gain a better understanding of the surface characteristics that influence power generation. For this, the zeta potential measurements were carried out to infer the surface charge. In the solution form, 1T' and 2H phases were found to have zeta potentials of -38.8 mV and -33.0 mV, respectively, which imply negative surface charges (Fig. S5). The negative surface charge on the 1T' phase arises collectively from the electron transfer from lithium to the MoS$_2$ crystal as well as S vacancies. In the 2H phase, Li is absent (as confirmed by the XPS studies), and hence, S vacancies predominantly contribute to its negative surface charge[23,24]. The contact angle revealed ≈34° for 1T' and ≈37° for 2H phase membranes (Fig. S6), indicating hydrophilicity, which ensures proper wetting of both samples. Similar hydrophilicity suggests that the interaction of water with the membrane surface is not very different in both types of samples; hence, the difference in power density is at least not related to hydrophilicity.

*Mechanism of energy generation:*

The mechanism of energy generation in the presence of an ambient environment lies in the intrinsic properties of the active material. In the absence of contact with liquid water, the moisture present in the ambient environment is responsible for generating power. However, this power is low due to the absence of a significant potential difference between the two electrodes. To investigate the nature of power generation, we focus on the reasons behind the enhanced power density in the presence of contact with liquid water.

MoS$_2$ possesses a negative surface charge, as indicated by the negative zeta potential value above. As the EEG is partially immersed in water (Fig. 5a), H$^+$ ions from the water (pH ~5.7 due to CO$_2$ dissolution in water) get adsorbed on the membrane surface, forming an EDL. The water rises in the channels due to capillary action. The H$^+$ ions move with the rising water and reach the top electrode. This creates a potential difference between the bottom and top electrodes. The external circuit connected to the electrodes measures the voltage generated by this streaming of ions, called streaming potential. As the water from the top portion of the membrane evaporates, it creates a water(H$^+$)-deficit condition, wherein the water flows upwards, creating a continuous capillary flow (Fig. 5b). Due to the negative surface charge of the 1T' MoS$_2$ channels, only the H$^+$ ions travel inside the channels, whereas any negative charges are repelled by the negatively charged channel walls and remain immobile. In contrast with the 1T' membrane, the 2H membrane lacks the presence of well-defined channels for the flow; therefore, the path for water intake is through surface water diffusion, i.e., across the channels. In the absence of liquid water, ambient moisture contains comparatively fewer H$^+$ ions to drive the energy. Thus, the power generated is lower in ambient moisture.

The higher powers obtained in the case of the 1T' device can be attributed to its structure and the presence of well-defined channels. However, these are absent in a 2H membrane, and the only route available for water is to diffuse across the flakes. Thus, surface water diffusion is the primary transport path, and it alone contributes to hydrovoltaic energy generation. The lamellar channels present in 1T' provide unhindered pathways for the transport of water. There is fast capillary action as well as evaporation from the surface of the open area of the membrane. Thus, in addition to the water diffusion across the flakes, the water transport along the channels also contributes towards the generated power in the 1T' phase. It can be concluded that for energy generation, there is a contribution from surfaces in the 2H phase and both surface and channels in the 1T' phase. Our EEG devices and measurements make it possible to clearly distinguish the surface and capillary effect, and this system can be used as a reference for such measurements.

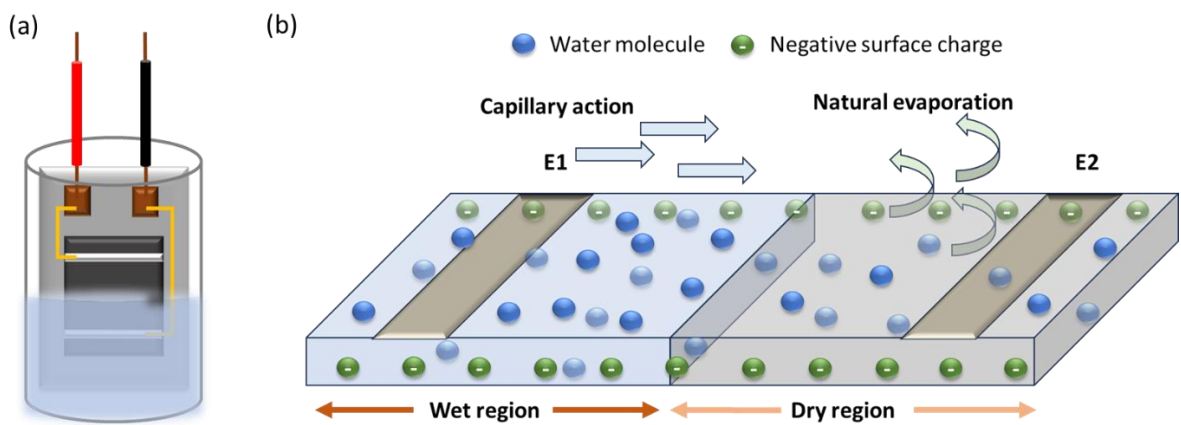

Fig. 5. Mechanism of energy generation. (a) EEG in contact with liquid water. (b) Schematic representation of capillary action and evaporation processes occurring continuously when the EEG is in contact with water.

To analyze the role of membrane functional groups on the generated power, we conducted a Fourier transform infrared (FTIR) study (Fig. S7). The FTIR spectra of the 2H phase show Mo-S and S-S stretching vibrations along with a low-intensity peak corresponding to -OH bond stretching vibration[28] at ~3332 cm$^{-1}$. Hence, this cannot be a source of negative surface charge on 2H MoS$_2$ laminates. On the other hand, the 1T' membrane shows Mo-S, S-S stretching vibrations, and a water bonding peak, but no functional groups were observed (Supplementary Section S2.4). In literature, the surface functional groups on the membranes have been reported to boost the output power[1,6,8]. FTIR study hence implies that energy generation in our membranes does not have any contribution from functional groups. Thus, it is reasonable to state that other factors, such as structure, channels, and conductivity, are more dominant in determining the output power.

*Streaming Potential and Current:*

For a deeper understanding of the energy generation mechanism in our EEGs, we conduct a theoretical analysis of the capillary pressure, streaming current, and potential. Let us consider a channel of length $L$, cross-sectional area $A$, and internal surface zeta potential ζ. The streaming current and potential due to external pressure difference ΔP are given by the following expressions[29,30]

$$I_S = \frac{A\, \varepsilon_0\, \varepsilon_r\, \Delta P\, \zeta}{\eta L} \ \ \ \ \ \ \ (1)$$

$$V_S = \frac{\varepsilon_0\, \varepsilon_r\, \Delta P\, \zeta}{\eta \sigma} \ \ \ \ \ \ \ \ (2)$$

where $\varepsilon_0$ is the vacuum permittivity, $\varepsilon_r$ dielectric constant of the liquid, $\sigma$ conductivity of the fluid-saturated porous medium, and $\eta$ liquid viscosity, respectively. These formulae were used to theoretically estimate the values of streaming current and potential (Supplementary Section S3).

The 1T' MoS$_2$ membrane consists of well-defined channels for the transport of water. $\Delta P$ is the capillary pressure inside the channel given by the Young-Laplace equation as $\Delta P = 4\gamma cos\theta/d$, where $\theta$ is the water contact angle, $\gamma$ is the surface tension of water, and $d$ is the diameter of the channel. Using $d$ as the h$_{vdW}$ calculated above as 6 Å, it was calculated to be ~4 kbar. Substituting this $\Delta P$ in eqn (1), $I_S$ was estimated to be 1.0 × 10$^{-6}$ A. Similarly, from eqn (2), $V_S$ was calculated to be 5.1 × 10$^{-4}$ V. The theoretical power density was 46.7 mW.m$^{-2}$. Our measured experimental values of current and voltage were $I_S$ ~10$^{-7}$ A and $V_S$ ~10$^{-4}$ V and power density was 2.0 mW.m$^{-2}$. We notice that the measured power is an order lower than the theoretically estimated value because of the lower value of the measured current after stabilization. When the water is added to the beaker, there is an instantaneous increase in the current level to ~3.5 μA (Fig. S3a), which then stabilizes to ~10$^{-7}$ A. With water as the dielectric between the MoS$_2$ flakes, a capacitive component arising from the initial wetting could have contributed to the initial increase in the measured current. We also observed a slight decrease in the XRD peak intensity of the 1T' phase upon exposure to liquid water, probably suggesting some conversion of the 1T' to 2H phase (Fig. S8a).

On the other hand, for a 2H MoS$_2$ membrane, surface water diffusion is the primary transport path due to the lack of well-defined channels. Thus, $\Delta P$ here is the saturated vapor pressure, 23 mbar. From eqn (1) and eqn (2), this leads to $I_S$ = 1.6 × $10^{-8}$ A and $V_S$ = 12.9 mV. Our measured experimental values of current and voltage were $I_S$ ~$10^{-8}$ A and $V_S$ ~10 mV. The theoretical power density was 5.8 µW.m$^{-2}$, close to the measured power density of 2.4 µW.m$^{-2}$. Thus, the theoretical values agree with our measured experimental values, and we can indeed say that the energy from our EEGs is generated by streaming current and potential.

Our estimations suggest that the conductivity of the 1T' phase is ~$10^3$ S m$^{-1}$ (Supplementary Section S3), very similar to previous reports[31,32] for the metallic 1T' phase. Similarly, the conductivity of the 2H phase comes out to be ~$10^{-4}$ S m$^{-1}$, also close to the conductivity reported earlier for the 2H phase[22]. The two phases have different structures with different electronic conductivities. As the metallic 1T' phase resulted in higher power density than the semiconducting 2H phase, it can be concluded that along with the structure, the electronic conductivity also significantly influences hydrovoltaic energy generation.

*Stability of MoS$_2$ phases:*

Having demonstrated hydrovoltaic energy generation from MoS$_2$ membranes and their theoretical estimations, we now discuss their feasibility in practical applications. The stability of the membrane is a crucial factor for this purpose. During our study, we noticed that the 1T' membrane starts to lose its 1T' phase in > 48-72 hours when stored in ambient laboratory conditions (Fig. S8b). The phase change process was delayed by keeping the membranes in a vacuum at room temperature. It could not be preserved for more than a week and is completely changed to the 2H phase after 10 days (Fig. S8c). On the other hand, 2H membranes showed no such instability under ambient conditions (Fig. S8d). Thus, for practical applications, the 1T' phase might not be a reasonable alternative unless its instability is resolved.

Some studies have reported strategies to stabilize the metallic phase. It has been suggested that the adsorption of monolayer water on both sides of the nanosheets could reduce the restacking, prevent their aggregation in water, and stabilize the metallic phase[33]. The 1T' phase is unstable because of the incomplete occupation of the degenerate orbitals. DFT calculations show that metal substrates, which can act as electron donors, can stabilize the metastable 1T' phase[34]. Similarly, covalent surface functionalization with an electron-donating functional group can also provide stability to this phase[21]. Supplementary Table 1 summarizes the important properties of both the phases.

## 3. Conclusions

In summary, we demonstrated the energy generation from ambient moisture and liquid water using two distinct phases of MoS$_2$, which have different conductivities and structures. We utilized 2D membranes and an uncomplicated geometry as opposed to complex device structures (like core-shell) used in other hydrovoltaic devices. We synthesized the metallic phase by chemical exfoliation, prepared membranes, and then converted them to the semiconducting phase by annealing in an inert atmosphere. Our experiments reveal that structure and channels play a significant role in water evaporation-based energy generation. Using a 1T' MoS$_2$ device (which has a higher electronic conductivity and large interlayer spacing than 2H), we were able to generate a power density of 2.0 mW.m$^{-2}$, three orders higher than that of 2H MoS$_2$. This study on MoS$_2$ can be used as a model system for probing hydrovoltaic mechanisms. We believe our study will contribute to the advancement of hydrovoltaic energy and direct further investigations in the field of sustainable energy generation.

**Experimental Section**

*Chemical exfoliation of MoS$_2$* - Lithium intercalation method was used to exfoliate MoS$_2$. For this, 5 mL of n-butyllithium (in 1.6 M hexane) (Avra Synthesis Private Limited) was added to 0.5 g MoS$_2$ powder (Asbury Graphite Mills, Inc) with moderate string under argon atmosphere for seven days. The mixture was washed thrice with hexane (Carbanio) to remove the excess organolithium reagents and organic byproducts. It was washed twice with water and centrifuged at 13500 rpm to remove inorganic byproducts and non-exfoliated particles.

*Preparation of MoS$_2$ membranes* - After centrifugation, the supernatant was the obtained MoS$_2$ dispersion, and had a yield of 0.6 mg/ml. It was used to prepare the 1T' MoS$_2$ membranes by vacuum filtration on a PVDF substrate (0.22 μm pore size, Merck). The membranes were then washed thoroughly with DI water before EEG device fabrication. These membranes were metallic 1T' in nature. The semiconducting 2H phase membrane was obtained by annealing the membranes at 200 °C in an inert atmosphere.

*Fabrication of the water evaporation-based electricity generator (EEG) device* - For EEG fabrication (Fig. 1b), a dry membrane was pasted on an acrylic sheet of a larger size. Silver was used as the electrode, one at the bottom of the strip (E1) and the other (E2) at a distance of 5 mm from E1. Gold wires were used to connect the silver electrodes to the printed circuit boards (PCBs) pasted on acrylic, and connecting wires were soldered for measurement. Keithley Source Meter 2636B was used to measure current and voltage across the electrodes of the EEGs.

**Data availability**

All data required to understand the conclusions in the paper are presented in the main text and supplementary information. Additional data related to this paper are available from the corresponding author upon request.

**Author contributions**

**Kalon Gopinadhan:** Conceptualization, Resources, Writing – Original Draft, Writing - Review & Editing, Supervision, Project administration, Funding acquisition **Kaushik Suvigya:** Methodology, Validation, Formal analysis, Investigation, Data curation, Writing – Original Draft, Writing – Review & Editing, Visualization, Project administration **Saini Lalita:** Formal analysis **Siva Nemala Sankar:** Formal analysis

**Andrea Capasso:** Formal analysis, Writing - Review & Editing **Li-Hsien Yeh:** Formal analysis, Writing - Review & Editing

## Conflict of interests

The authors declare no competing interests.

## Acknowledgments

This work was mainly funded by DST-INAE with grant no. 2023/IN-TW/09. We also acknowledge the financial support from MHRD STARS with grant no. MoE-STARS/STARS-1/405 and also by Science and Engineering Research Board (SERB), Government of India, through grant no. CRG/2019/002702. K. G. acknowledges the support of Kanchan and Harilal Doshi chair fund. The authors acknowledge the contribution from the IITGN central instrumentation facility and Common Research & Technology Development Hub on Chemical Processes. L. H. Y. acknowledges the financial support from the National Science and Technology Council (NSTC), Taiwan under Grant No. NSTC 112-2923-E-011-003-MY3.

# Supplementary Information

**Contents:**

**Section S1: Materials and methods**

**Section S1: Materials and methods**

*S1.1 Materials*

$MoS_2$ bulk powder (Asbury Graphite Mills, Inc.), n-hexane and n-Butyllithium (Carbanio), and distilled water (Millipore water purification system of conductivity 18.2 MΩ) were utilized for the study. All reagents were used as such without any further purification. PVDF supporting membranes (pore size = 0.22 µm) were purchased from Merck.

*S1.2 Physical characterization and sample preparation*

Surface morphology and elemental analysis of the $MoS_2$ nanosheets were done using FE-SEM, JSM7600F with Oxford energy dispersive X-ray spectroscopy (EDX) attachment. The as-prepared membranes were analyzed in SEM in surface and cross-sectional modes. For cross-sectional analysis, the membrane was torn to expose the layers. The crystalline structure of prepared $MoS_2$ nanosheets was determined by X-ray diffraction studies using an automated multipurpose X-ray diffractometer by Rigaku SmartLab at 1.54 Å Cu Kα. XRD was done on a whole membrane or its smaller part. Raman measurements were performed with an ALPHA300 R Confocal Raman Microscope (WITec) using 532 nm laser (0.8 mW power) for excitation at room temperature. XPS spectra were acquired with an ESCALAB 250 XI (Thermo Fisher Scientific, Source: Al Kα 1486.6 eV, 650 µm spot size, Pass energy: 40 eV with hemispherical analyzer) system with an analysis chamber maintained in ultrahigh vacuum (UHV ~ $5 \times 10^{-10}$ mbar) conditions. The functional groups in the membranes were analyzed by using a Fourier transform infrared (FTIR) spectrometer. Samples for ultraviolet-visible spectroscopy and zeta potential measurement were prepared by diluting the master solution with DI water in a ratio of 1:10.

**Section S2: Figures**

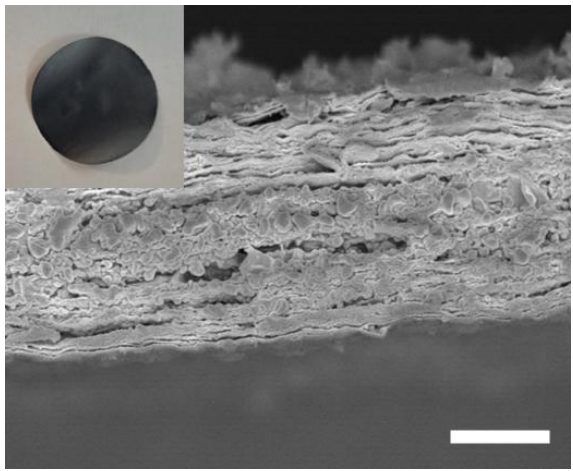

**Fig. S1. Cross-sectional SEM image of a 2H MoS$_2$ membrane.** Scale bar 500 nm. Inset: camera image of a freestanding membrane. Although the cleaving of the membrane is not proper, the laminate structure is still observed clearly.

*S2.1 van der Waals gap in MoS₂ phases*

In the context of 2D materials, van der Waals distance and van der Waals gap are essentially different terms. The interlayer space, also called $d$ spacing, is the space between two consecutive $MoS_2$ layers. This is calculated from the XRD analysis using the Bragg's law, $2\,d\sin\theta = n\lambda$. This $d$ is called the van der Waals distance ($d_{vdW}$). However, not all of this interlayer space is available for transport, as some space is also occupied by the electron cloud. The free space that is available for the transport of a species is called van der Waals gap ($h_{vdW}$). This is calculated by subtracting the thickness of one layer of $MoS_2$, i.e. 6.5 Å., from the $d$ spacing.

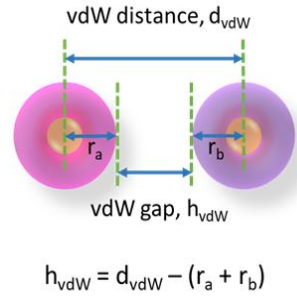

$$h_{vdW} = d_{vdW} - (r_a + r_b)$$

**Fig. S2. Schematic representation of van der Waals gap.**

For 2H phase, from XRD data, we have $d_{vdW}$ (2H $MoS_2$) = 6.5 Å.

∴ $h_{vdW}$ (2H $MoS_2$) = 6.5 Å - 6.5 Å = 0.

The van der Waals gap in 2H $MoS_2$ is negligible.

For 2H $MoS_2$, $d_{vdW}$ = 6.5 Å. Due to its compact structure, we take 6.5 Å as the upper bound to estimate the $h_{vdW}$ in 1T' $MoS_2$ as follows:

$$h_{vdW} = d_{vdW} - (r_a + r_b)$$

From XRD data, we calculated $d_{vdW}$ (1T' $MoS_2$) = 12.5 Å,

And let $r_a + r_b$ = 6.5 Å,

∴ $h_{vdW}$ (1T' $MoS_2$) = 12.5 Å - 6.5 Å = 6 Å

Hence, $h_{vdW}$ for 1T' $MoS_2$ is 6 Å.

*S2.2 XPS Characterization*

X-ray photoelectron spectroscopy (XPS) was used for phase confirmation of the synthesized 1T' $MoS_2$ and 2H $MoS_2$. The peaks of 1T' phase are downshifted by ≈0.5 eV compared to the 2H phase. In the Li-intercalated 1T' phase, the Mo 3d core level peaks chemically shift such that its binding energies are lower than that of 2H phase. The decrease in binding energy is due to the chemical reduction of Mo from the +4 to the +3-oxidation state.

*S vacancies:* Ideally, in a defect free $MoS_2$, the ratio S:Mo is 2:1. In our case, for 1T' phase membrane, S:Mo = 1.32:1, whereas for 2H phase membrane, S: Mo = 1.35:1. This is summarized in the following table:

| **Case** | **Ideal** | **1T'** | **2H** |
|---|---|---|---|
| **S:Mo** Ratio | 2:1 | 1.32:1 | 1.35:1 |

From XPS, we can infer that S vacancies are indeed present in both our phases. In fact, 1T' phase has higher number of S vacancies than the 2H phase.

Please note that the XPS data for both the phases was taken after a week of the sample preparation, as the samples were in transit to another institute for analysis. This might have slightly affected the composition of the 1T' $MoS_2$ membrane.

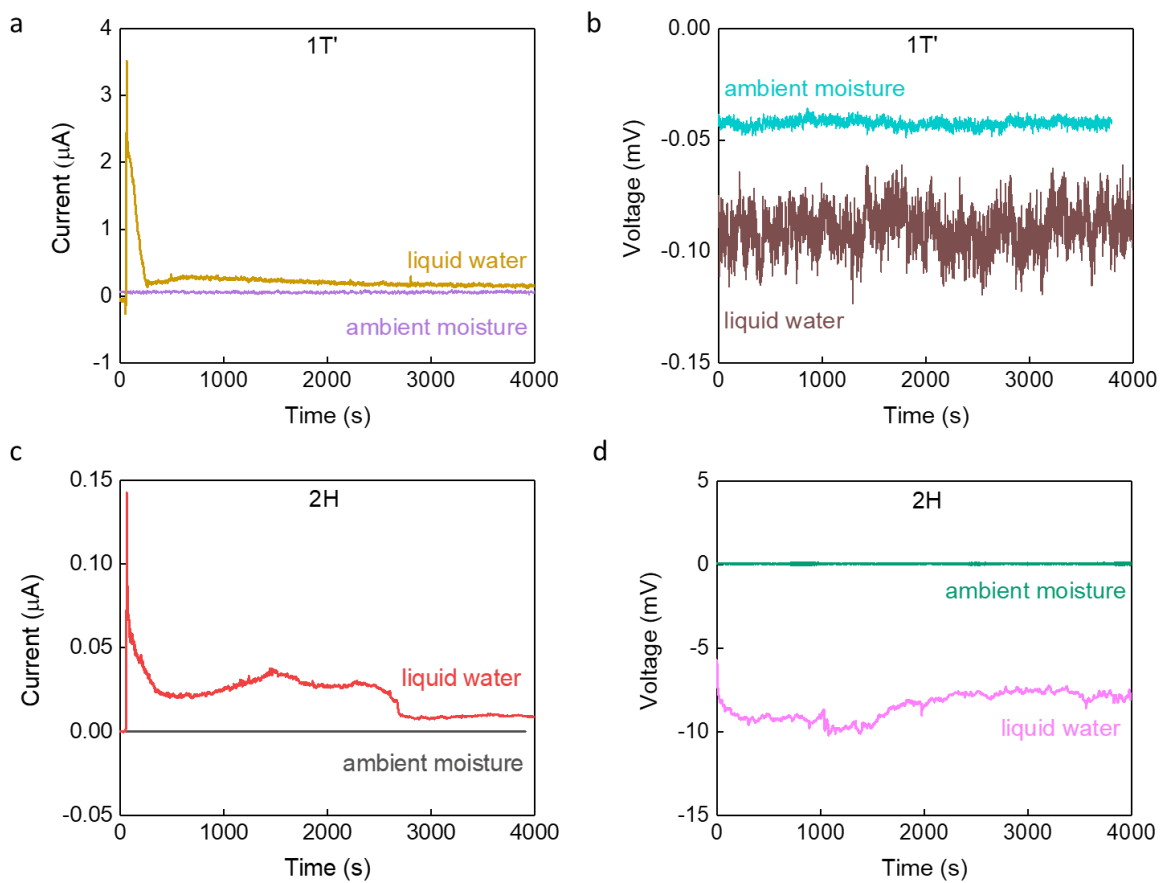

**Fig. S3. Variation of current and potential with time. a. and b.** 1T' devices. **c. and d.** 2H devices. When the water is added to the beaker, both devices have an instantaneous increase in the current level, which can be seen from the spike just after t = 0.

*S2.3 Additional experiments*

To prove conclusively that the streaming current is in fact driven by evaporation, we further demonstrate that there is a strong dependency on the environmental conditions such as humidity and air flow, by performing additional experiments with a 1T' device as follows:

a) Firstly, during the measurement, the EEG was sealed properly using parafilm, and it was observed that the current level gradually dropped. When the device was unsealed again, the current regained its value. This was repeated a few times for different time intervals.
b) Secondly, air flow was turned on above the EEG, and it was observed that the current level increased. This is because the water evaporates faster in the wind. Such change could be completely repeated.

These experiments confirm that evaporation is the origin of the induced current.

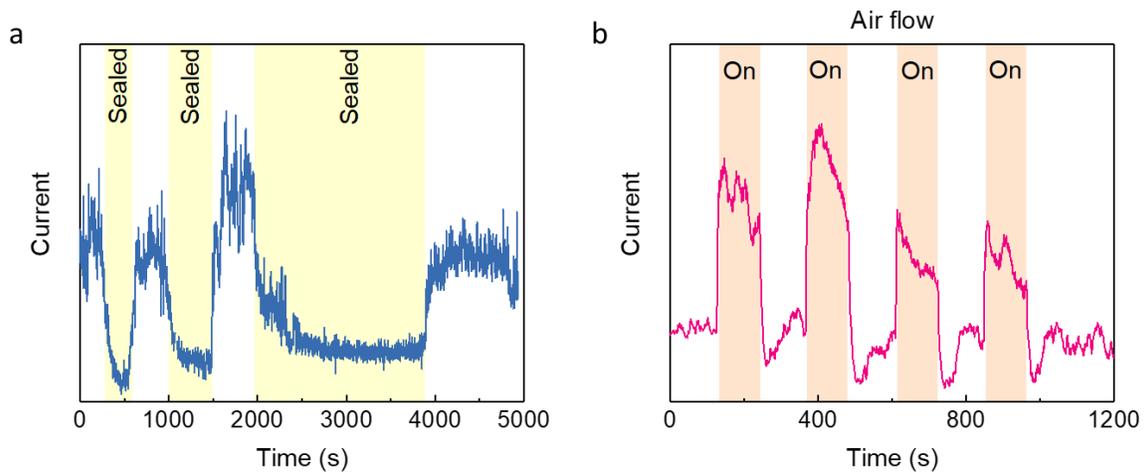

**Fig. S4. Experiments to determine the effect of change in evaporation rate. a.** The EEG was sealed inside the beaker, and then unsealed. The dropping of current by 90% inside the sealed device is due to slower evaporation. **b.** The air flow was turned on and off above the EEG. The current increased by 12 times with the air flow due to faster evaporation.

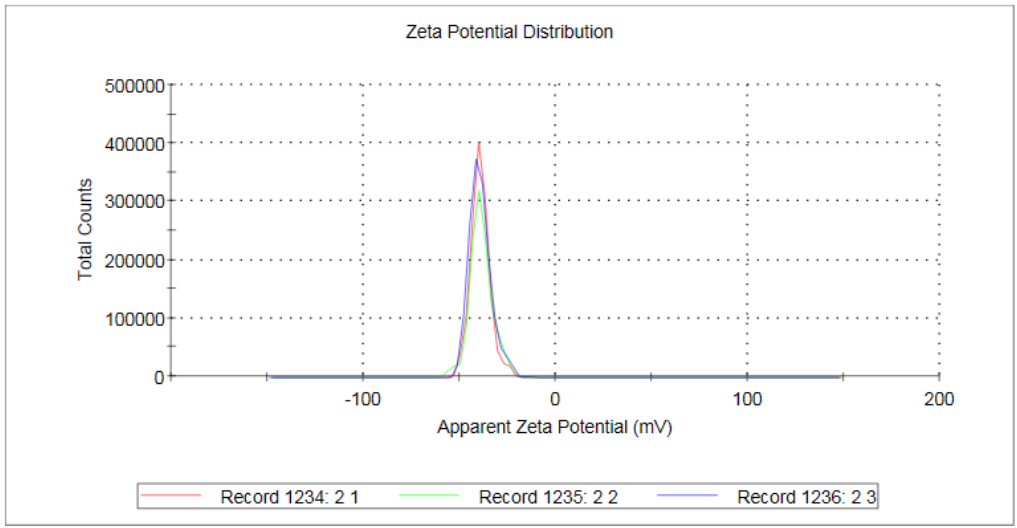

**Fig. S5. Zeta Potential of 1T' MoS$_2$ dispersion.** Zeta Potential measurements of the provided a mean value of -38.8 mV. The different curves are from three different measurements.

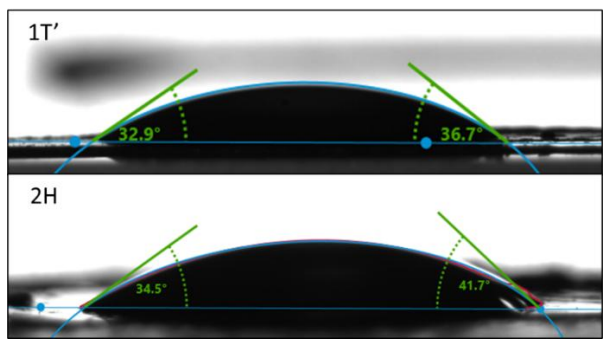

**Fig. S6. Contact angle measurement of 1T' and 2H MoS$_2$ membranes.** The contact angle for 1T' phase ≈34° and for 2H phase ≈37°.

*S2.4 FTIR Analysis*

1T' and 2H phases membranes were analyzed using FTIR spectrometer. The 463 cm$^{-1}$ peak in 1T' phase corresponds to the Mo-S bond. The 1T' phase shows a water bonding peak[1] at 1600 cm$^{-1}$ which indicates the presence of water trapped between its interlayers, as also indicated by the XRD data. This peak is absent in the 2H phase, whereas it shows other peaks, such as S-S bond[2] at 808 cm$^{-1}$, and -OH bond[3] at ~3332 cm$^{-1}$.

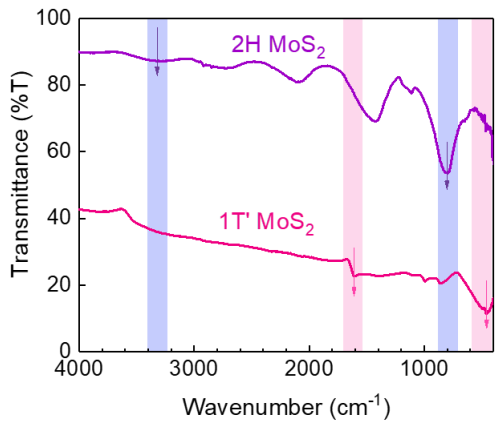

**Fig. S7. FTIR analyses of two MoS$_2$ phases**

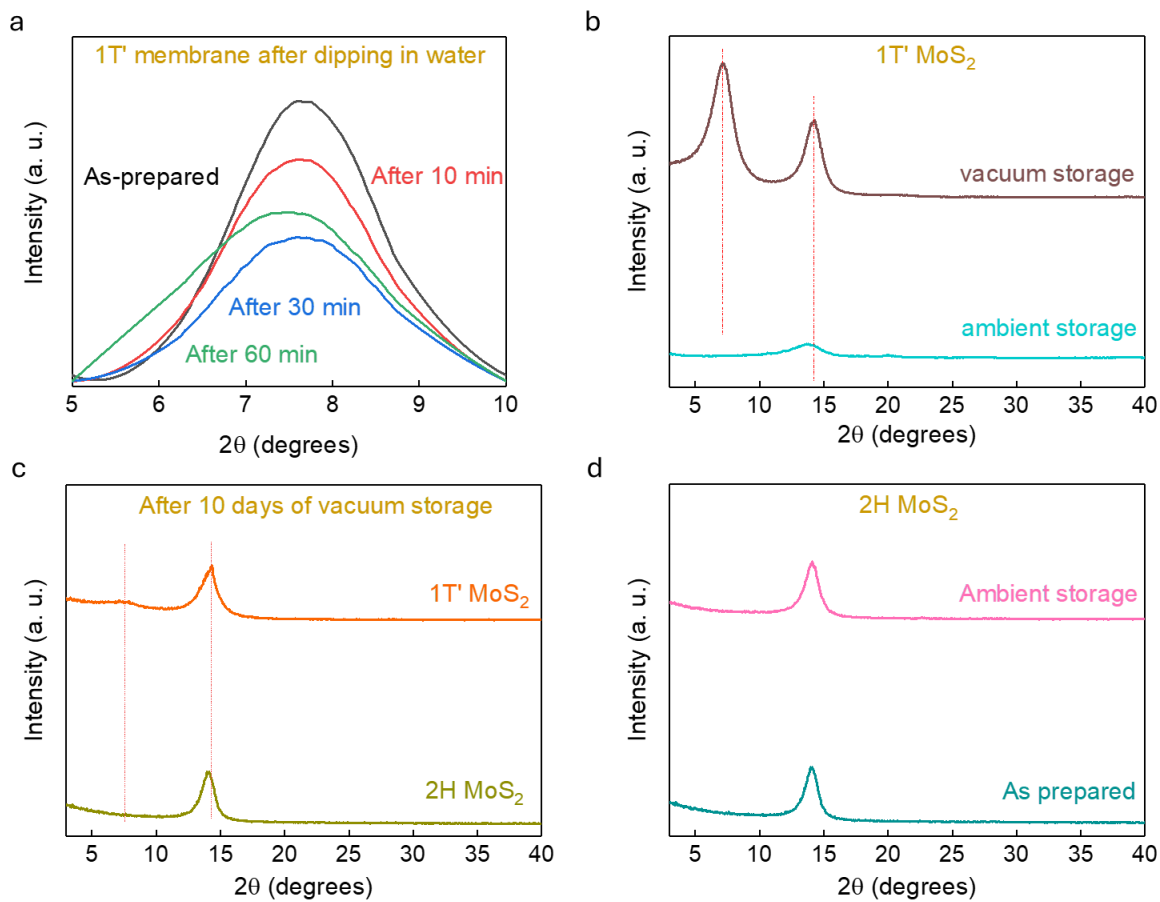

**Fig. S8. X-ray diffraction data for analyzing the stability of 1T' MoS$_2$ and 2H MoS$_2$. a.** 1T' phase in contact with water. **b.** Effect of storage condition on the metastable 1T' phase- the 1T' phase is lost if the membrane is stored in the ambient environment. **c.** Decay of 1T' phase over time, even after storage in a vacuum. **d.** Effect of storage condition on the 2H phase- no change is observed over a long time.

**Section S3: Theoretical estimation of Streaming Current and Potential**

The streaming current and potential due to external pressure difference $\Delta P$ are given by:

$$I_S = \frac{A \, \varepsilon_0 \, \varepsilon_r \, \Delta P \, \xi}{\eta L} \quad ....(1)$$

$$V_S = \frac{\varepsilon_0 \, \varepsilon_r \, \Delta P \, \xi}{\eta \sigma} ....(2)$$

where $L$ is channel length, $A$ is the pore cross-sectional area and $\zeta$ is the internal surface zeta potential, $\varepsilon_0$ is the vacuum permittivity, $\varepsilon_r$ dielectric constant of the liquid, $\sigma$ conductivity of the fluid-saturated porous medium, and $\eta$ liquid viscosity, respectively.

The channel length for the calculations is taken as $L$ = distance between electrodes E1 and E2 = 5 mm, and the thickness of the membranes, t = 1 µm.

- *For a 1T' MoS$_2$ membrane:*

$\Delta P$ is the capillary pressure given by $\Delta P = 4\gamma \cos\theta/d$, where $\theta$ is water contact angle, $\gamma$ is the surface tension of water, and $d$ is the diameter of the channel (capillary/ pore).
Considering $\gamma$ = 0.072 N.m$^{-1}$, $\theta$ = 34°, and $d$ = 6 Å (the h$_{vdW}$ of 1T' phase),

$\Delta P = 4\gamma \cos\theta/d$ = 3.98 kbar = ~4 kbar

Pore cross-sectional area was calculated as $A = \pi d^2/4$ = 2.82 × 10$^{-19}$. The width of 1T' membrane was 11 mm. Now, $\varepsilon_0$ = 8.854 × 10$^{-12}$ F m$^{-1}$, $\varepsilon_r$ = 3[4,5], $\eta$ = 0.89 mPa.s.

Total no. of channels, n = $\frac{t \times w}{A}$ = 3.9 × 10$^{10}$.

The conductivity $\sigma$ of 1T' phase was estimated from the slope of the I-V graph as 9.09 × 10$^2$ S m$^{-1}$ = ~10$^3$ S m$^{-1}$.

Eqn (1) would give the current for one channel, so it must be multiplied with the total number of channels responsible for water intake. This could be calculated as follows:

Substituting the above values in eqn (1) × n,

$I_S = 1.0 \times 10^{-6}$ A.

The factor n has already been considered while calculating $\sigma$, so from eqn (2),

$V_S = 5.1 \times 10^{-4}$ V.

Power density = $\frac{I_S \times V_S}{A}$ = 46.7 mW m$^{-2}$

- *For a 2H MoS$_2$ membrane:*

The $\sigma$ was estimated from the slope of the I-V graph as $1.8 \times 10^{-4}$ S m$^{-1}$. Due to lack of well-defined channels,

    (i) $A$ is the area between the two electrodes E1 and E2. $A = L \times w$ (w = width of the membrane) = 5 mm × 7 mm.

    (ii) $\Delta P$ is the saturated vapor pressure = 23 mbar.

Substituting these values in eqn (1) and eqn (2),

$I_S = 1.6 \times 10^{-8}$ A, and

$V_S = 12.9$ mV.

Power density = $\frac{I_S \times V_S}{A}$ = $5.8 \times 10^{-3}$ mW m$^{-2}$ = 5.8 µW.m$^{-2}$

- *For a purely surface contribution in 1T' phase:*

The $\sigma$ can be estimated from slope of the I-V graph as 0.18 S m$^{-1}$.

    (i) $A$ is the area between the two electrodes E1 and E2. $A = L \times w$ (w = width of the membrane) = 5 mm × 11 mm.

    (ii) $\Delta P$ is the saturated vapor pressure = 23 mbar.

Substituting these values in eqn (1) and eqn (2),

$I_S$(surface, 1T') = $3.0 \times 10^{-8}$ A, and

$V_S$ (surface, 1T') = $1.5 \times 10^{-5}$ V.

Power density (surface, 1T') = $\frac{I_S \times V_S}{A}$ = 7.8 nW m$^{-2}$

Since the surface contribution is three orders of magnitude less than the channel contribution, it would not make a significant difference to the overall values of current and power density.

**Supplementary Table 1. Comparison between 1T' and 2H phases of MoS$_2$.**

| Characterization/Properties | 1T' | 2H |
|---|---|---|
| Phase | metallic | semiconducting |
| Thermal stability | metastable | stable |
| Channel structure | well-defined channels | no channels |
| Interlayer spacing | 12.5 Å | 6.5 Å |
| van der Waals gap | 6 Å | - |
| Contact angle | ≈34° (hydrophilic) | ≈37° (hydrophilic) |
| Zeta Potential | -38.8 mV | -33 mV |
| Power density in contact with water | 2.0 mW.m$^{-2}$ | 2.4 µW.m$^{-2}$ |
| Power density in ambient | 0.1 mW.m$^{-2}$ | 0.2 pW.m$^{-2}$ |
| Path of water transport | Surface water diffusion (across the flakes) and along the channels | Surface water diffusion (across the flakes) only |